\documentclass{emulateapj}
\newcommand{\e}{\epsilon}

\newcommand{\epthr}{\epsilon_{thr}^\prime}
\newcommand{\epkp}{\epsilon_{pk}^\prime}

\newcommand{\ep}{\epsilon^\prime}
\newcommand{\gp}{\gamma_p^\prime}
\newcommand{\g}{\gamma}
\newcommand{\xp}{\gamma_p^{\prime 2}}
\newcommand{\tp}{t^\prime}

\newcommand{\pp}{p^\prime}

\newcommand{\Np}{N^\prime}
\newcommand{\up}{U^\prime}

\begin{document}

\title{Rapid X-ray Declines and Plateaus in Swift 
GRB Light Curves 
Explained by A Highly Radiative Blast Wave}

\author{Charles D. Dermer}

\affil{E. O. Hulburt Center for Space Research, Code 7653
Naval Research Laboratory, Washington, D.C. 20375-5352}

\email{ dermer@gamma.nrl.navy.mil}

\begin{abstract}
GRB X-ray light curves display rapid declines followed by 
a gradual steepening or plateau phase in 
$\gtrsim 30$\% of GRBs in the Swift sample.
Treating the standard relativistic blastwave model in a uniform circumburst medium, it
is shown that if GRBs accelerate ultra-high energy cosmic rays through a
Fermi mechanism, then the hadronic component can be rapidly depleted
by means of photopion processes on time scales $\sim 10^2$ -- $10^4$ s
after the GRB explosion.  While discharging the hadronic energy in the
form of ultra-high energy cosmic ray neutrals and escaping cosmic-ray
ions, the blast wave goes through a strongly radiative phase, causing 
the steep declines observed with Swift. Following the discharge,
the blast wave recovers its adiabatic behavior, forming the observed
plateaus or slow declines. These effects are illustrated by calculations
of model bolometric light curves. The results show that steep X-ray declines and
plateau features occur when GRB sources take place in rather 
dense media, with $n \gtrsim 10^2$ cm$^{-3}$ out to $\gtrsim 10^{17}$ cm.
\end{abstract}


\section{Introduction}

The Swift telescope is providing a new database of GRB light
curves consisting of a BAT light curve in the 15 -- 150 keV range
followed, after slewing within $\approx 100 $ s, by a detailed 0.3 -- 10
keV XRT X-ray light curve and UVOT monitoring \citep{geh04}.
Extrapolating the BAT light curve to the XRT range gives $\sim 1$ keV
X-ray light curves since the trigger.  \citet{obr06} present a catalog
of the combined 0.3 -- 10 keV light curves of 40 GRBs, of which $ 14$
or so have measured redshifts, $\approx 30$\% display rapid X-ray
declines, and an additional $\approx 30$\% display features unlike
simple blast wave model predictions. In some GRBs, 
the 0.3 -- 10 keV flux can decrease by 4 or 5 orders of magnitude over
a period of  $\lesssim 10^2$ seconds within several minutes after the GRB trigger
(e.g., GRB 050915B, GRB 050422, GRB 050819).
About one-half the XRT sample shows X-ray flares or short timescale 
($\Delta t/t\ll 1$) structure at $\gtrsim 10^3$ s after the GRB trigger,
and in some cases out to $\gtrsim 10^5$ s (e.g., GRB 050904 at $z = 6.29$
and GRB 050730 at $z = 3.97$).

Following the rapid X-ray declines, a more gradual steepening commences
in most XRT light curves, as in the cases of GRB 050219A or GRB 050607.
In several GRBs, e.g., GRB050315 or GRB050822, a plateau phase
with rising and decaying features within $0.1$ -- $ 1$ day after the 
GRB trigger are observed;
this phase may also be found in other GRBs (GRB 050724 or GRB 050819),
but could be contaminated or mimicked by a late-time X-ray flare. In 
a few GRBs (e.g., GRB 050401, GRB 050717), the X-ray decline is gentle 
and monotonic, 
but in general the XRT light curves, whether displaying overall convexity or concavity, 
reveal temporal structure and, oftentimes, X-ray flares (e.g., GRB 050712, GRB 050716, GRB 050726). 
A phenomenological model with two distinct components can fit 
the rapid decays and hardenings \citep{wil06}, but lacks a physical basis. 
An acceptable physical model for GRB afterglows must be able to 
explain this diversity of behaviors.

A combined leptonic-hadronic GRB model is proposed in this
paper as the cause of the rapid X-ray declines and plateaus discovered
with Swift. The analysis presented here 
follows the standard blastwave model \citep[e.g.,][]{mes06}. We show
that if GRBs accelerate cosmic rays to ultra-high energies, then for
certain classes of GRBs, the blastwave will become strongly
radiative during the early afterglow, and consequently will exhibit rapid
X-ray declines. This class of GRBs is defined by a range of blast wave
and environmental parameters. One set of parameters that
produces such declines has initial Lorentz factors $\Gamma_0
\sim 100$ -- 300, apparent total energy releases $\gtrsim 10^{54}$
ergs (and absolute energy releases $\lesssim 10^{52}$ ergs), 
and surrounding medium density $n \gtrsim 10^2$ cm$^{-3}$,
which is assumed to be proton dominated.  The blast wave microphysical
parameters $\e_e, \e_B$ must both be $\gtrsim 0.1$. 

For these GRBs, Fermi processes in the blast wave are assumed to
accelerate proton and ions, like they do electrons, to
ultrarelativistic energies. By making reasonable approximations for the 
acceleration rate as a fraction of the Larmor rate, and particle
escape through Bohm diffusion, we find that 
 photohadronic losses and particle escape significantly deplete the internal
energy of the blast wave, causing the blast wave dynamics to be strongly affected.
Photopion interactions by the
ultrarelativistic protons and secondary neutrons with the internal synchrotron photons make a
source of escaping neutrons, neutrinos and cascade
$\gamma$ rays, in addition to a generally weaker proton synchrotron 
component. 

In Section 2, a standard blast-wave physics
analysis for emissions from a GRB external shock in the 
prompt and early afterglow phase is presented, and timescales for the various processes
are calculated for an adiabatic blast wave that decelerates by
sweeping up material from a uniform surrounding medium. 
Parameter sets that allow a large fraction of the internal 
energy to be radiatively discharged through
hadronic processes
are graphically examined in Section 3. In Section 4, the equations for blastwave evolution are
solved in the case of internal energy that is promptly radiated or 
is exponentially depleted with time. 
Synthetic bolometric light curves are calculated. Although these cannot be directly
compared to the Swift X-ray light curves without taking into account 
spectral effects and more complicated radiation physics, the bolometric light curves 
exhibit many of the features observed with Swift. 

Discussion of multiwavelength and multichannel  $\gamma$-ray, 
cosmic-ray, and neutrino predictions for this model is found in Section 5, 
including a comparison with other models for the X-ray declines. The
study is summarized in Section 6.

\section{Analysis}

The energy flux $\Phi_E = {dE/ dA dt} = \nu F_\nu = f_\e = {L_*/ 4\pi
d_L^2},$ where $d_L=10^{28}d_{28}$ cm is the luminosity distance, 
the source luminosity $L_* = 4\pi x^2 c u_0^\prime \Gamma^2$, 
$x$ is the distance of the blast wave from the explosion center,
and $\Gamma = \Gamma(x)$ is the blast wave Lorentz factor at $x$. The measured
dimensionless photon energy $\epsilon \equiv h\nu/m_e c^2$ is related
to the emitted photon energy $\ep$ through the relation
$\ep \cong {(1+z)\e/ 2\Gamma}$, and the differential distance traveled
by the blast wave during reception time $dt$ is  $dx \cong {2\Gamma^2 c dt/ (1+z)}$, 
where $z$ is the source redshift, and primes denote
quantities in the comoving fluid frame. Thus the comoving energy
density 
\begin{equation}
u^\prime _{\ep} = m_e c^2 \e^{\prime 2}n^\prime_{ph}(\ep )
\cong\; {k_{kin}\; {d_L^2 f_\e \over  c x^2 \Gamma^2}}\;,
\label{kkin}
\end{equation}
where $k_{kin} \approx 1/3$ gives
the photon energy density that produces a received
spectrum with peak $\nu F_\nu$ flux $= f_\e$ \citep{der04}. 
Relating $x$ to the measured variability time gives the proper 
spectral density of the radiating fluid as a function of the Doppler factor.
The GRB
spectrum is approximated by the broken power-law form
\begin{equation}
n^\prime_{ph}(\ep ) = {k_{kin} d_L^2 f_{\e_{pk}} \over 
c x^2 \Gamma^2 m_ec^2}\; {[u^a H(1-u) + u^b H(u-1)]\over \e^{\prime~2} }\,,
\label{nprime}
\end{equation}
where $u = \e/\e_{pk} = \ep/\ep_{pk}$, $f_{\e_{pk}} = 10^{-6} f_{-6}$
ergs cm$^{-2}$ s$^{-1}$ is the peak $\nu F_\nu$ flux at the $\nu
F_\nu$ peak photon energy $\e_{pk}$, $a > 0 $ and $b < 0$ are the $\nu
F_\nu$ indices, and the Heaviside function $H(y) = 1$ for $y \geq 1$ and
$H(y) = 0$ otherwise restricts the lower
and upper branches of the spectrum to their respective ranges.

\subsection{Adiabatic Blast Wave}

Consider a blast wave with coasting Lorentz factor $\Gamma_{300} =
\Gamma_0/300$ and apparent {\it total} isotropic energy release $E_0 =
10^{54}E_{54}$ ergs, so that its absolute energy release, 
due to collimation of the relativistic outflows, 
is $\lesssim 10^{52}$ ergs. If the blast wave sweeps through a uniform 
surrounding medium with
proton density $n_0 = 100 n_2$ cm$^{-3}$, it will slow down on the
deceleration length scale
\begin{equation}
x_d = {3E_0\over 4\pi m_pc^2 n_0\Gamma_0^2}\cong 2.6\times 10^{16}
\big( {E_{54}\over n_2 \Gamma_{300}^2}\big)^{1/3}\;{\rm cm}\;
\label{xd}
\end{equation}
\citep{mr93}. A relativistic adiabatic blast wave decelerates
 according to the relation
\begin{equation}
\Gamma = \Gamma_0 /\sqrt{1+2(x/x_d)^3}\;
\label{Gamma}
\end{equation}
\citep{bd00}, from which can be derived the asymptotes
\begin{equation}
{x\over x_d} =
\cases{ 
 \tau, &$\tau \ll 1$\cr\cr
(2\tau)^{1/4}, &$\tau \gg 1$\cr}\;
\label{xxd}
\end{equation}
and
\begin{equation}
{\Gamma\over \Gamma_0} = 
\cases{ 
 1, &$\tau \ll 1$\cr\cr
2^{-7/8}\tau^{-3/8}, &$\tau \gg 1$\cr}\; .
\label{GoverG0}
\end{equation}
Here the dimensionless time $\tau \equiv t/t_d = t^\prime/t_d^\prime$, 
where the deceleration timescale is 
\begin{equation}
t_d = {(1+z) x_d \over \Gamma_0^2 c} \cong 
9.6(1+z)\big( {E_{54}\over n_2 \Gamma_{300}^8}\big)^{1/3}\;{\rm s}\;,
\label{td}
\end{equation}
and the inverse of the comoving deceleration timescale is
\begin{equation}
t_d^{\prime -1}= ({x_d \over \Gamma_0 c})^{-1} 
\cong 3.5\times 10^{-4} \big( {n_2 \Gamma_{300}^5\over E_{54}}\big)^{1/3}
\;{\rm s}^{-1}\;.
\label{tdp}
\end{equation}
The available time in the comoving frame is
\begin{equation}
\tp_{ava} = t_d^\prime
\cases{ 
 \tau, &$\tau \ll 1$\cr\cr
(2^{17/8}/5)\tau^{5/8}\cong 0.872\tau^{5/8}, &$\tau \gg 1$\cr}\; ,
\label{tpava}
\end{equation}

\subsection{Blast Wave Physics}

We treat the photopion process in the fast cooling regime
\citep{spn98}.  The minimum Lorentz factor $\gamma_{min} = \e_e {\cal F}_p
m_p\Gamma/m_e$, where ${\cal F}_p = (p-2)/(p-1)$ and $2 < p < 3$.  The
emission detected by Swift is assumed to be predominately nonthermal synchrotron
radiation. The mean magnetic field in the fluid frame is $B = b B_{cr}
= \sqrt{32 \pi m_pc^2 n_0\epsilon_B } \Gamma\cong 0.4 \sqrt{\e_B
n_0}\;\Gamma{\rm~G}
\cong 370 \sqrt{\e_{B-1}n_2}\;\Gamma_{300}(\Gamma/\Gamma_0){\rm~G}$, 
where $B_{cr} = m^2_ec^3/e\hbar \cong 4.414\times 10^{13}$ G, and the
minimum mean observed synchrotron photon energy from
electrons with $\gamma_{min}$ is
\begin{equation}
\e_{min} \cong {\Gamma b \gamma_{min}^2 \over 1+z} 
\cong {0.85 \sqrt{\e_{B-1} n_2}\e_{e-1}^2 {\cal F}_{5/2}^2 
\Gamma_{300}^4(\Gamma/\Gamma_0)^4\over 1+z}\;.
\label{emin}
\end{equation}

The cooling Lorentz factor $\gamma_c = 3m_e(1+z)/16 m_p \sigma_{\rm T}
\e_B n_0 \Gamma^3 c t_d\tau$, and the dimensionless cooling frequency 
(in units of $m_ec^2$) is given by
\begin{equation}
\e_{c}
\cong {2.3\times 10^5 (1+z) \over (\e_B n_0)^{3/2}
\Gamma^4 t_d^2 \tau^2 }\;\cong\; {1.0\times 10^{-8} 
\Gamma_{300}^{4/3}(\Gamma/\Gamma_0)^{-4}\over 
\e_{B-1}^{3/2} n_2^{5/6} (1+z) E_{54}^{2/3}\tau^{2}}\;.
\label{ec}
\end{equation}
Comparing eqs.\ (\ref{emin}) and (\ref{ec}) shows that we are in the
strong cooling regime, $\epsilon_c < \epsilon_{min}$, when
\begin{equation}
\tau \lesssim  7\times 10^5\;(\e_{e-1}\e_{B-1}{\cal F}_{5/2})^2 n_2^{4/3}  
E_{54}^{2/3}\Gamma_{300}^{8/3} \;.
\label{tau}
\end{equation}

The photon energy at the peak of the
$\nu F_\nu$ synchrotron spectrum 
for the fast-cooling blast wave is given by 
\begin{equation}
\e_{pk} \cong\;{0.85 \sqrt{\e_{B-1}n_2}\e_{e-1}^2{\cal F}_{5/2} 
\Gamma_{300}^4\over 
1+z}
\cases{ 
 1, &$\tau \ll 1$\cr\cr
0.09 \tau^{-3/2}, &$\tau \gg 1$\cr}\;,
\label{epk11}
\end{equation}
and the $\nu F_\nu$ peak flux is given by
\begin{equation}
f_{\e_{pk}}\cong {\Gamma^2\over 4\pi d_L^2}\;
({4\over 3}c\sigma_{\rm T}U_B)\; \gamma_{min}^3 N_e^\prime(\g_{min};x)\;.
\label{fepk1}
\end{equation}
In this expression, $U_B = B^2/8\pi$ is the magnetic-field energy density in the
comoving frame. For the fast-cooling regime, $N_e^\prime (\g_{min})
\cong N_e(x)\g_c\g_{min}^{-2}$, and $N_e(x) = 4\pi n_0 x^3/3$. Thus
$$f_{\e_{pk}}(10^{-6}~{\rm ergs~cm^{-2}~s^{-1}}) \cong$$
\begin{equation}
 2.7\;{\e_{e-1} {\cal F}_{5/2} n_2^{1/3} E_{54}^{2/3} \Gamma_{300}^{8/3}\over
d_{28}^2}
\cases{ 
 \tau^2, &$\tau \lesssim  1$\cr\cr
(2\tau)^{-1}, &$\tau \gg 1$\cr}\; .
\label{fepk2}
\end{equation}

\subsection{Photopion Losses}

The energy-loss rate due to photopion production on the GRB synchrotron
radiation field is 
$$r_{\phi\pi} = t^{\prime -1}_{\phi\pi} ( \gp
)\;\cong \;{c\over 2\xp }
\int_0^\infty d \ep \;{n^\prime_{ph}(\ep )\over 
\e^{\prime 2}}\;\times$$
$$\int_0^{2\gp\ep} d\ep_r \;\ep_r 
\sigma^K_{\phi\pi}(\ep_r ) \approx 
{k_{kin} \hat\sigma d_L^2 f_{\e_{pk}} \over x^2 
\Gamma^2 m_ec^2 \e_{pk}}\;\times
$$
$$\{ H(1-y) \big[{y^2\over 3-a} - {2y^{a-1}\over (3-a)(a-1)}
 + {1\over a-1}\big]
$$
\begin{equation}
+ \;\big[{\max(1,y)^{b-1}\over 1-b} - 
{y^2 \max (1,y)^{b-3}\over 3-b}\big]\}\;,
\label{tprime_1}
\end{equation}
after substituting eq.\ (\ref{nprime}) into eq.\ (\ref{tprime_1}), with $y \equiv
\epthr/2\gp\epkp$. Here we use the approximation of \citet{ad03},
where the product of the photopion cross section and inelasticity is
$\sigma^K_{\phi\pi}(\ep_r ) = \hat\sigma \cong 70~ \mu$b $H(\ep_r
-\epthr)$, and the threshold dimensionless photon energy for photopion 
production is $ \epthr \cong 400$ (i.e., $m_ec^2 \epthr\approx 200$ MeV).

The asymptotes of eq.\ (\ref{tprime_1}) are
\begin{equation}
 t^{\prime -1}_{\phi\pi}(\gamma_p) \cong K_{\phi\pi}(\bar\gamma_p)\;
\cases{ 
 y^{b-1}, &$y\geq 1$\cr\cr
{(3-b)(a-b)\over 2(a-1)}\;,&$y\ll 1,\;1 \leq a \leq 3$\cr\cr
{(1-b)(3-b)\over (3-a)(1-a)}\;y^{a-1}, &$y\ll 1,\;0 \leq a \leq 1$\cr}\; 
\label{tphipp_1}
\end{equation}
where $y = 1$ defines the Lorentz factor $\bar\gamma_p^\prime$ of
protons that interact primarily with internal synchrotron photons at
the $\nu F_\nu$ peak frequency $\epkp$. We call 
$\bar E_p = m_pc^2 \bar\gamma_p = m_pc^2
\Gamma\bar\gamma_p^\prime$ the {\it peak
cosmic-ray proton energy}, as it is the characteristic energy of
protons with Lorentz factor $\bar\gamma_p^\prime$ 
that would escape from the blast wave with Lorentz $\bar
\gamma_p$ factor as measured by a local observer. Hence the 
peak cosmic-ray proton energy is 
\begin{equation}
\bar E_p = {m_p c^2 \Gamma^2 \epthr
\over (1+z) \epkp}\cong {1.7\times 10^{16} 
(\Gamma/300)^2\over ({1+z\over 2}) \e_{pk}}\;{\rm eV}\;,
\label{barep}
\end{equation} 
and
$$K_{\phi\pi}(\bar\gamma_p)\;
 = {2k_{kin}\hat\sigma d_L^2 f_{\e_{pk}}
\over x^2\Gamma^2 m_ec^2 \epkp (1-b)(3-b) }\cong$$
\begin{equation}
 {1.1\times 10^{-6} k_{kin} d_{28}^2 f_{-6}\over 
(1-b)(3-b)(1+z) x_{16}^2 (\Gamma/300)\e_{pk}}\;{\rm s}^{-1}\;,
\label{Kphipi}
\end{equation}
where $x = 10^{16}x_{16}$ cm.

\subsection{Rates and Limits for Ultrarelativistic Protons}

\subsubsection{Adiabatic Loss Rate}

Adiabatic expansion losses operate
on the same timescale as the available time, so for comparison 
of the adiabatic
loss rate to other rates, we write $t^{\prime -1}_{adi}
\cong t^{\prime -1}_{ava}$, using eq.\ (\ref{tpava}).

\subsubsection{Photopion Loss Rate}

Substituting eqs.\ (\ref{xxd}), (\ref{GoverG0}), 
and (\ref{epk11}) into eq.\ (\ref{tphipp_1})
gives the comoving photopion energy-loss rate
\begin{equation}
 t^{\prime -1}_{\phi\pi}(\bar E_p) 
\cong {5.2\times 10^{-7}k_{kin} n_2^{1/2}\;{\rm s}^{-1}\over
(1-b)(3-b) \Gamma_{300}^2 \e_{B-1}^{1/2}\e_{e-1} }
\cases{ 
 1, &$\tau \lesssim 1$\cr\cr
2^{23/8}\tau^{3/8}, &$\tau \gg 1$\cr}\;
\label{tphipp_2}
\end{equation}
at the characteristic energy 
\begin{equation}
\bar E_p ({\rm eV}) \cong {4\times 10^{16}\over 
\sqrt{\e_{B-1}n_2}\e_{e-1}^2 {\cal F}_{5/2}\Gamma_{300}^2}
\cases{ 
 1, &$\tau \lesssim 1$\cr\cr
3.4\tau^{3/4}, &$\tau \gg 1$\cr}\;
\label{barEp}
\end{equation}
of an escaping cosmic ray, as measured by a stationary 
observer in the local source frame. 

\subsubsection{Acceleration Rate}

Because a significant energy gain by a particle can take place through
Fermi acceleration mechanisms on times not shorter than the Larmor
time $t^\prime_{\rm L} = mc^2 \gp/eBc = (mc/eB)(\gamma/\Gamma)$
\citep{rm98}, the acceleration rate in the proper frame can be
written as $r_{acc} = t^{\prime -1}_{acc} = \zeta_{acc}t^{\prime-1}_{\rm
L},$ with the acceleration parameter 
$\zeta_{acc}\lesssim 1$.  Hence the acceleration rate at the
peak cosmic-ray proton energy $\bar E_p$ is given by
$$r_{acc}(\bar E_p) \;\cong $$
\begin{equation}
 25\zeta_{acc}\;{n_2 \e_{B-1}
\Gamma_{300}^4 \e_{e-1}^2 {\cal F}_{5/2}\;{\rm s}^{-1}
}\cases{ 
 1, &$\tau \lesssim 1$\cr\cr
{1\over 11.4\tau^{3/2}}, &$\tau \gg 1$\cr}\;
\label{tp_1acc}
\end{equation}
The acceleration rate for $10^{20}E_{20}$ eV cosmic ray protons is
$$r_{acc}(E_{20})   \cong $$
\begin{equation}
{1.1\times 10^{-3} (\zeta_{acc}/0.1)
\sqrt{\e_{B-1} n_2} 
\over  \, E_{20} }{\Gamma_{300}^2\over 
(1+ 2^{7/8}\tau^{3/8})^2}\;\;{\rm s}^{-1}\;.
\label{tp_1acc1}
\end{equation}
Here we consider a standard acceleration parameter $\zeta_{acc} \cong 0.1$. 
Values of $\zeta_{acc}\cong 1$ require unreasonably efficient 
particle acceleration, with particles gaining a large fraction of their
energy in a single Larmor timescale. If $\zeta_{acc} \ll 0.1$, then GRBs
would not accelerate cosmic rays sufficiently rapidly to make ultra-high
energy cosmic rays.

\subsubsection{Escape Rate}

The mean escape rate using the Bohm diffusion approximation is given
by $\tp_{esc} = \langle x \rangle^2/2\kappa_{\rm B}$, where
$\kappa_{\rm B} = c^2\tp_{\rm L}/3$ is the diffusion coefficient. 
For particle acceleration in GRB blast waves, the
characteristic dimension $\langle x \rangle $ is the shell width
$\Delta^\prime = f_{\Delta}x/\Gamma$, and $f_\Delta \cong 1/12$ for
the width of the shocked fluid shell swept up from the circumburst medium
by an adiabatic relativistic blast waves \citep[e.g.,][]{pm99}. Thus
the escape rate is
\begin{equation}
t^{\prime -1}_{esc} \cong {2c E_p \Gamma \zeta_{esc}\over 3eB f_\Delta^2 x_d^2}\;
{1\over (x/x_d)^2}\;, 
\label{tp_1}
\end{equation}
where $\zeta_{esc}$ is a parameter that allows particle escape on 
timescales shorter ($\zeta_{esc} >1$) or longer ($\zeta_{esc} < 1$)
than the escape timescale set by Bohm diffusion. If the GRB blast-wave
shell entrains a randomly oriented, tangled magnetic field,
then depending on the coherence length of the disordered 
magnetic field, the particles could diffuse more rapidly than
given by Bohm diffusion, so that $\zeta_{esc} \gtrsim 1$. In contrast,
if the GRB blast wave is assumed to entrain an
ordered field, for example, a toroidal geometry in a jetted 
fireball, then escape could be impeded compared to the Bohm
timescale, so that $\zeta_{esc} \ll 1$.  Here we consider the Bohm 
limit for the rate at which particles escape, keeping in mind that the actual escape
rate could be quite different.

The escape rate for protons with characteristic energy $\bar E_p$,
eq.\ (\ref{barEp}), is
\begin{equation}
t^{\prime -1}_{esc}(\bar E_p ) \cong {5\times 10^{-7}\zeta_{esc}\;{\rm s}^{-1}
\over \e_{B-1} n_2^{1/3} \e_{e-1}^2 {\cal F}_{5/2}\Gamma_{300}^{2/3}}
\;\cases{ 
 \tau^{-2}, &$\tau \lesssim 1$\cr\cr
2.4 \tau^{1/4}, &$\tau \gg 1$\cr}\;
\;.
\label{tp_2}
\end{equation}
The escape rate for $10^{20}E_{20}$ eV cosmic ray protons is
$$t^{\prime -1}_{esc}( E_{20} ) \cong$$
\begin{equation}
 {1\times 10^{-3}E_{20}  n_2^{1/6}
\Gamma_{300}^{4/3}\zeta_{esc}\;{\rm s}^{-1}\over \sqrt{\e_{B-1}} E_{54}^{2/3}}
\;\cases{ 
 \tau^{-2}, &$\tau \lesssim 1$\cr\cr
(2\tau)^{-1/2}, &$\tau \gg 1$\cr}\;.
\label{tp_3}
\end{equation}

\subsubsection{Size Scale Limitation}

We also have the \citet{hil84} condition that the Larmor radius be 
smaller than the characteristic size scale of the system, which is
the shell width $\Delta^\prime$. Requiring $r_{\rm L} = (mc^2/ eB) (\gamma/\Gamma )
 \lesssim \Delta^\prime \cong
{f_\Delta x}/\Gamma$ implies a limit to maximum proton energy, given by
\begin{equation} 
E^{\rm H}_p(10^{20}{\rm~eV}) = 2.4\; \e^{1/2}_{B-1} n_2^{1/6} 
\Gamma_{300}^{1/3}
E_{54}^{1/3}\;\cases{ 
 \tau, &$\tau \lesssim 1$\cr\cr
 {\tau^{-1/8}\over 2^{5/8}}, &$\tau \gg 1$\cr}\;,
\label{hillas}
\end{equation}
using the asymptotes, eqs.\ (\ref{xxd}) and (\ref{GoverG0}). Note the
slow late-time decline $ E^{\rm H}_p \propto t^{-1/8}$ when $\tau \gg
1$ \citep{vie98,bd98}.  Thus we see that standard parameter values
allow Fermi acceleration of protons to ultra-high energies in GRB
blast waves when $\zeta_{acc} \gtrsim 0.1$ and 
$\zeta_{esc} \cong 1$, making GRBs a viable candidate for UHECR production.

\subsubsection{Proton Synchrotron Energy Loss Rate}

The inverse of the synchrotron energy-loss timescale for an escaping
proton with Lorentz factor $\gamma_p = \Gamma\gamma^\prime_p$ is 
$$r_{p,syn} = t^{\prime
-1}_{p,syn} (\gamma_p) = |{\dot\gamma^\prime_{p,syn}\over\gp} | ={16\over
3}\;c\sigma_{\rm T} n_0 \e_B \Gamma \;\big({m_e\over m_p}\big)^2
\gamma_p $$
\begin{equation}
\cong 9.4\times 10^{-6} n_2 \e_{B-1}\Gamma_{300} E_{20} 
\;\big({\Gamma\over 300}\big)\;{\rm s}^{-1}\;.
\label{tprimepsyn}
\end{equation}
The mean proton synchrotron photon energy, in units of $m_ec^2$,
 from protons with
energies $10^{20} E_{20}$ eV ($E_{20} \cong \gamma_{11} = \gamma_p/10^{11}$) 
as measured in the stationary frame, is given
by
\begin{equation}
\e_{p,syn} = {3\Gamma\over 1+z}\;{B\over B_{cr}}\;
{m_e\over m_p}\;\gamma_p^{\prime 2}\;
\cong 4.5\times 10^5\;{\e_{B-1} n_2 E_{20}^2\over 1+z}\;,
\label{epsyn}
\end{equation}
independent of time in the relativistic deceleration phase---provided of course that $\e_B$ is time-independent
and $n_2 =const.$

\section{Results}

For initial parameter estimation, we adopt the Standard Parameter Set 
given in Table 1, with $z = 1, \Gamma_{300} =
1, E_{54} = 1, n_2 = 1, \e_{e-1} = 1,$ and $\e_{B-1} = 1$. 
This set is
motivated by values that reproduce typical peak fluxes and durations
for BATSE GRBs at $z \approx 1$ \citep{cd99}, except that here $\e_B =
0.1$ rather than $\e_B \cong 10^{-4}$. 
%We also keep standard here and 
%in the following the values $\zeta_{acc} = 0.1$
%and $\zeta_{esc} = 1$. 

%\clearpage
\begin{deluxetable}{cccc}

\tablecaption{
Parameter Sets Used in Analysis\label{table1}}
\tablehead{    &
\multicolumn{1}{c}{Stnd.\  Set} & 
 Set 1&
 Set 2
}
\startdata
%\tableline
%\begin{tabular}{cddd}
\tableline
\tableline
$z$ & 	 1 & 	1 & 	 1 \\	
$\Gamma_{300}$ & 	 1 & 	1 & 	0.5   \\	
$E_{54}\tablenotemark{a}$ & 	 1 & 	1 & 	 10   		 \\
$n_2\tablenotemark{b}$ & 	 1 &  10 & 	 10  	 \\
$\e_{e -1}$ & 	 1 &   3 & 	 1  	\\
$\e_{B -1}$ & 	 1 & 3 & 	 3  \\	
$\zeta_{acc}$ & 	 0.1 & 	0.1 &  0.1  \\	
$\zeta_{esc}$ & 	 1 & 	1 & 	 1  \\      
\tableline	
$t_d$(s) & 	19.3 & 8.9 & 122  \\      
$x_d$($10^{16}$ cm) & 	2.6 & 1.2 & 4.1   \\  
\tableline
\enddata
%\end{tabular}
\tablenotetext{a}{Apparent isotropic energy release in units of $10^{54}$ ergs.}
\tablenotetext{b}{Circumburst medium density, assumed uniform and dominated
by H, in units of 100 cm$^{-3}$.}
\end{deluxetable}

%\clearpage
\begin{figure}[t]
\epsscale{0.80}
\plotone{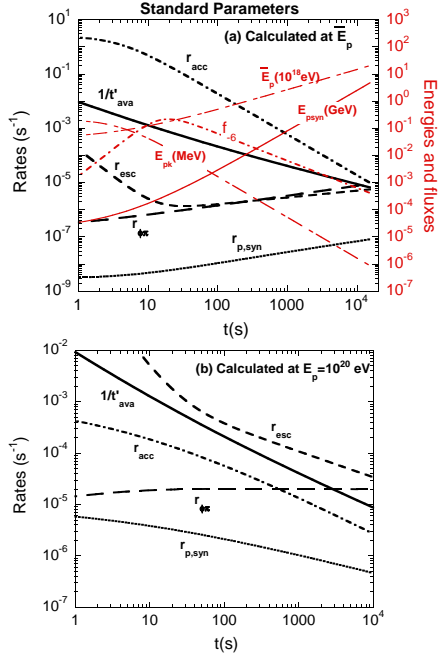}
%\vskip-0.25in
\caption{(a) Characteristic rates and energies calculated at $\bar
E_p$, (b) characteristic rates calculated at $E_p = 10^{20}$ eV, using
the Standard Parameter set.  }
%\vskip0.2in
\label{f1}
\end{figure}
%\clearpage

Fig.\ 1(a) shows the Standard Parameter rates of acceleration,
photopion losses, proton synchrotron losses and escape for cosmic rays
with energy $\bar E_p$, the $\nu F_\nu$ peak photon energy $\e_{pk}$,
the cosmic-ray peak energy $\bar E_p$, and the mean proton synchrotron
photon energy $E_{p,syn} = m_ec^2\e_{p,syn}$ radiated by protons with energy $\bar E_p$.
Here and throughout we use an acceleration factor $\zeta_{acc} = 0.1$,
an escape factor $\zeta_{esc} = 1$,
and kinematic factor $k_{kin} = 1$.  

From the top panel in Fig.\ 1, one sees that the acceleration rate exceeds the
inverse of available time throughout the early afterglow phase, so
cosmic rays with energies $\sim\bar E_p$ are in principle easily
accelerated through Fermi processes to energies $\gtrsim\bar E_p$. Only at several hours into the
afterglow do photopion losses limit acceleration to $E_p
\lesssim \bar E_p$, which by then is $\gtrsim 10^{19}$ eV.  
At these late times, proton synchrotron emissions
make a $\gtrsim $1\% contribution to the total loss rate.  The diffusive
escape rate of protons can appear as a $\gtrsim $1\% effect on the
total rate, but is generally insignificant in the early
afterglow. The external shock emission from this GRB is 
brightest $\approx 20$ s after first being detected, though shell
collisions could make brighter features during the prompt 
and afterglow phases (see Discussion).

The acceleration, escape, and loss rates for a cosmic ray proton with
$E_p = 10^{20}$ eV are plotted in Fig.\ 1b. As can be seen, there
isn't enough time to accelerate cosmic rays to $\gtrsim 10^{20}$ eV
energies for these parameters, so there cannot be significant $\gtrsim
10^{20}$ eV (super-GZK) cosmic ray production or photopion losses from
such GRBs (unless $\zeta_{acc} \gg 0.1$). Protons would also escape before a significant fraction
could be accelerated to such energies.

%\clearpage
\begin{figure}[t]
\epsscale{0.8}
\plotone{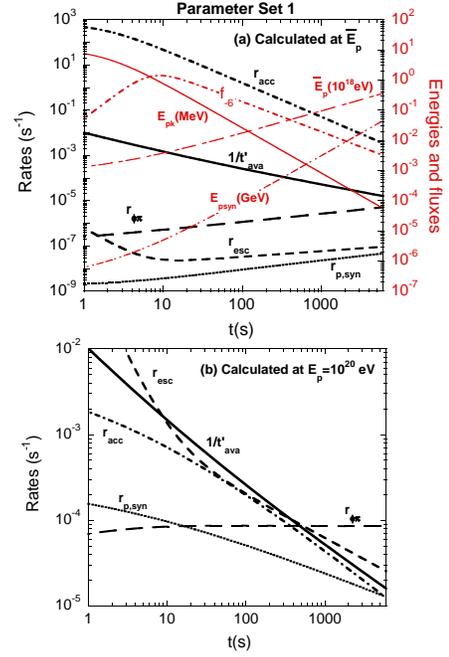}
\caption{As in Fig.\ 1, but calculated for Parameter Set 1, with the
energy of electrons, protons, and magnetic field in approximate
equipartition.  }
\label{f2}
\end{figure}
%\clearpage

A set of parameters that overcomes these limitations is easily
found. Consider Parameter Set 1 in Table 1, with $z = 1, \Gamma_0 = 300, E_{54} =
1, n_2 = 10, \e_{e-1} = 3, \e_{B-1} = 3,$ giving the rates, fluxes,
and energies shown in Fig.\ 2.  For $10^{20}$ eV cosmic rays shown in
the lower panel, an interesting conjunction occurs when $r_{\phi\pi}
\approx 1/\tp_{ava} \lesssim r_{acc}$, which happens here at $\approx
300$ s. Protons accelerated to $\approx 10^{20}$ eV energies are
converted, $\sim 1/2$ of the time\footnote{The charge-changing fraction 
is not $1/3$, as expected for $\Delta$ resonance excitation and 
decay, due to the inclusion of direct pion channels above threshold, 
and multi-pion production at energies far above threshold.}, to ultrarelativistic neutrons that
escape from the blast wave to form one component of a neutral beam
\citep{ad03}, in addition to neutrinos and $\gamma$ rays.

The GRB formed in Parameter Set 1, Fig.\ 2, has a rather high $E_{pk}\sim 8$ MeV---which
may be irrelevant in an internal/external scenario---but 
values of $ \Gamma_0\lesssim 300$ will lower $\e_{pk}$ during the
prompt phase and lengthen the prompt phase duration. Fig.\ 3 shows the
results for Parameter Set 2 with $z = 1, \Gamma_0 = 150, E_0 = 10^{55} {\rm~ergs},
n_2 = 10, \e_{e-1} = 1, \e_{B-1} = 3\;.$ This GRB peaks $\gtrsim 50$
s after the trigger, has a lower $\e_{pk}$, and reaches a slightly
lower $f_{\e_{pk}}$ peak flux than in Fig.\ 2. 

Parameter Sets 1 and 2 model fast-cooling GRBs with $\e_e = 0.3$ and $\e_e = 0.1$,
respectively, that exhibit a
radiative photopion phase.  By letting $\e_e$= 30\%  
for Set 1, it is understood that a large
fraction of the swept-up power is found in nonthermal
electrons rather than in baryons or fields.  A large body of parameter
values clustering around the Parameter Set 1 values predict strong
photopion losses in the early afterglow phase, especially if
$\zeta_{acc} \gtrsim 0.1$ is allowed.  If $\e_e \lesssim
0.1$ is assumed, as in Parameter Set 2, 
then agreement with GRB energetics for GRBs with {\it apparent bolometric $\gamma$-ray
energy} $ \gtrsim 10^{54}$ 
ergs, which represents a significant fraction of pre-Swift GRBs \citep{fb05}, 
means that a large fraction of GRBs also have {\it apparent total energy} $E_{54} \gtrsim 10$, 
due to the unknown efficiency of converting the total energy into $\gamma$ rays.  
Even for small $\e_e$ models,the necessary large values of
total apparent total energy can result in strong photopion losses. Thus if GRBs accelerate
UHECRs, then photohadronic losses 
efficiently deplete GRB internal 
energy and affect the blastwave dynamics over a wide range
of GRB parameters.

%\clearpage
\begin{figure}[t]
\epsscale{0.8}
\plotone{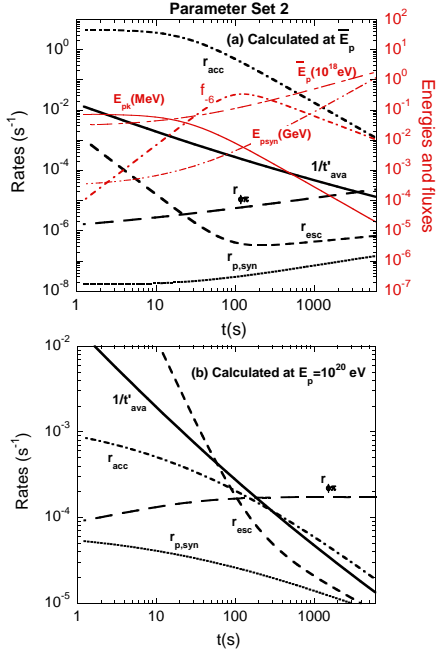}
%\vskip-0.1in
\caption{As in Fig.\ 1, but calculated for Parameter Set 2.  }
%\vskip0.1in
\label{f3}
\end{figure}
%\clearpage

\section{Blast Wave Evolution with Radiative Discharge }

The previous section showed that with reasonable parameter
values, photopion production and escape can rapidly deplete
internal blastwave energy in the early afterglow when
hadronic processes become important. We now calculate model light curves
resulting from the radiative discharge of internal energy 
through photohadronic processes. 
The dynamics of the blast wave is numerically solving
using the equation of relativistic blastwave evolution, 
including adiabatic losses, given by 
\begin{equation}
-\;{dP/dx\over \Gamma^2 - 1} = {\Gamma P \;{dm\over dx} + \big({\Gamma^2\over P}\big) 
{d\up_{adi}\over dx}\over M_0+m(x)+\up}\;
 \label{dPdx}
\end{equation}
\citep{dh01}. Here the blast wave momentum $P = \sqrt{\Gamma^2 -1}$, $M_0 =
E_0/\Gamma_0$ is the initial baryon loading, that is, the 
baryonic mass mixed into the explosion,
and $m(x)$ is the swept-up mass (all masses are now in energy units). 
The internal energy, excluding rest mass energy, 
is given by 
$$\up = \up(x) =  m_p\int_0^\infty dp^\prime (\gamma_p^\prime -1) \Np (\pp;x) \label{Utot}=$$
\begin{equation}
 4\pi m_p \int_0^x dx_i \; x_i^2 \;n_{0}(x_i) \;(\bar\gamma - 1)\;,
\label{U}
\end{equation}
where  $\pp = \sqrt{\gamma_p^{\prime 2} -1}$ is the proton's dimensionless momentum in the 
comoving frame, $\gamma_p^\prime$ 
its Lorentz factor, and
\begin{equation}
\bar p =  \bar p(x,x_i)
= \sqrt{\bar\gamma^2-1} = P(x_i)\;\big( {x_i\over x}\big)\;\big[{\Gamma(x)\over \Gamma(x_i)}\big]^{1/3}
\label{pbar}
\end{equation}
solves the equation for momentum evolution of a nonthermal proton in a blast wave with 
thickness $\Delta^\prime = f_\Delta x/\Gamma$ \citep{pm99,dh01}.
Here only adiabatic losses are considered for the protons, and  $f_\Delta $ is 
assumed to remain constant with $x$ and take the value $1/12$.
The
differential swept-up mass 
\begin{equation}
{dm(x)\over dx} = 4\pi m_p x^2 \;n_{0}( x) \;,
\label{dmxdx}
\end{equation}
where $n_{0}(x)$ is the circumburst medium density, assumed to be radially
symmetric about the GRB source.
The internal energy $\up$ changes due to volume expansion and adiabatic losses according to
the relation
\begin{equation}
{d\up_{adi}\over dx} = -\; {4\pi m_p\over x}\;(1 - 
{1\over 3}{d\ln \Gamma\over d\ln x}) \;\int_0^x dx_i\;x_i^2\;n_{0}(x_i) 
\big({\bar p^2\over \bar \g}\big)\;.
\label{dUadidx}
\end{equation}

We consider two case scenarios to simulate the change in internal energy due to a photohadronic discharge:
\begin{enumerate}
\item A fraction $\xi$ of the internal energy is instantaneously radiated away at $x = x_0$. 
Following the discharge, the blast wave is again assumed to sweep up circumburst material
and evolve adiabatically. In this case, $\up(x)$ and $d\up_{adi}/dx$ are given 
by eqs.\ (\ref{U}) and (\ref{dUadidx}), 
respectively, at $x\leq x_0$. At $x>x_0$,
%$${U(x)\over 4\pi m_p} = (1-\xi )\int_0^{x_0} dx_i \; x_i^2 \;n_{0}(x_i) \;(\bar\gamma - 1)+$$
$${\up(x)\over 4\pi m_p} = (1-\xi )\int_0^{x_0} dx_i \; x_i^2 \;n_{0}(x_i) \;(\bar\gamma - 1)+$$
\begin{equation}
\int_{x_0}^x dx_i \; x_i^2 \;n_{0}(x_i) \;(\bar\gamma - 1)\;,
\label{U(x)1}
\end{equation}
and
$$-{d\up_{adi}\over dx} = {4\pi m_p\over x}\;(1 - 
{1\over 3}{d\ln \Gamma\over d\ln x})\;\times$$
%$$\big[(1-\xi )\;\int_0^{x_0} dx_i\;x_i^2\;n_{0}(x_i) \big({\bar p^2\over \bar \g}\big) +$$
\begin{equation}
 \big[(1-\xi )\;\int_0^{x_0} dx_i\;x_i^2\;n_{0}(x_i) 
\big({\bar p^2\over \bar \g}\big) + \int_{x_0}^x dx_i\;x_i^2\;n_{0}(x_i) 
\big({\bar p^2\over \bar \g}\big)\;\big] .
\label{dUadidx7}
\end{equation}
\item The internal energy is radiated on an exponential dissipation timescale 
$\tp_{int}$, followed by a recovery of the blast wave to adiabatic behavior at $x\geq x_0$. Thus
$${\up(x)\over 4\pi m_p} =  \int_0^{x} dx_i \; x_i^2 n_{0}(x_i) (\bar\gamma - 1)
\exp[-\tp(x_i)/\tp_{int}]\; +$$
\begin{equation}
H(x-x_{0})\int_{x_{0}}^x dx_i \; x_i^2 \;n_{0}(x_i) \;(\bar\gamma - 1)\,
,
\label{U(x)2}
\end{equation}
and
$$-{d\up_{adi}\over dx} = {4\pi m_p\over x}\;(1 - 
{1\over 3}{d\ln \Gamma\over d\ln x})\;\times$$
%$$\big[\int_0^{x} dx_i\;x_i^2\;n_{0}(x_i) \big({\bar p^2\over \bar \g}\big)\exp[-\tp(x_i)/\tp_{int}]\; +$$
$$\big[\int_0^{x} dx_i\;x_i^2\;n_{0}(x_i) 
\big({\bar p^2\over \bar \g}\big)\exp[-\tp(x_i)/\tp_{int}]\;$$
\begin{equation}
  + H(x-x_{0})\int_{x_0}^x dx_i\;x_i^2\;n_{0}(x_i) 
\big({\bar p^2\over \bar \g}\big)\big] \;,
\label{dUadidx1}
\end{equation}
for all $x>0$.
\end{enumerate}

%\clearpage
\begin{figure}[t]
%\vskip-0.5in
\epsscale{0.9}
\plotone{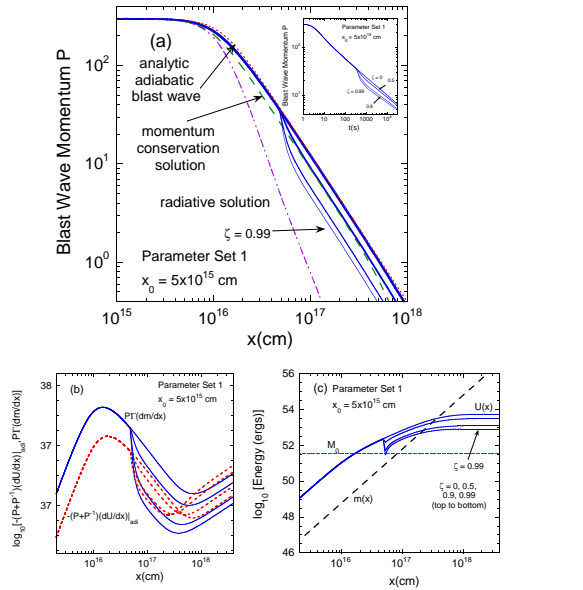}
%\vskip-0.8in
\caption{(a) Solid curves show the evolution of a GRB blast wave as a function of $x$ for different
fractional discharges $\xi = 0, 0.5, 0.9,$ and 0.99, for Parameter Set 1. 
Also shown are approximations to the case with $\xi = 0$  (see text). 
Inset shows blastwave evolution as 
a function of observer time. (b) Variation of values of the terms in the numerator 
of eq.\ (\ref{dPdx}), as labeled.  (c) Variation of values of the terms in the denominator 
of eq.\ (\ref{dPdx}), as labeled. }
%\vskip0.1in
\label{f4}
\end{figure}
%\clearpage

We consider first Case 1 with Parameter Set 1,\footnote{The microphysical parameters are irrelevant
in this treatment of blastwave dynamics, though this may not be true in general, particularly for radiative GRBs
with $\e_{B-1} > 1$  
 in the fast cooling regime.} with a sudden loss of radiative energy 
taking place at $x_0 = 5 \times 10^{16}$ cm, corresponding to an observer time
of $\approx 650$ s after the start of the GRB---essentially the same timescale
when photopion losses and escape become important (Fig.\ 2b).  
In Fig.\ 4(a), the numerical solutions to the equations for the evolution of 
$P $ are plotted for $\xi = 0, 0.5, 0.9, 0.99$. Also shown are some approximations
to the dynamics of the blast wave when $\xi = 0$ \citep[see][]{dh01}; the curve labeled ``momentum conservation solution" 
has the term $d\up_{adi}/dx $ set equal to zero; the curve
labeled ``analytic blast wave" is eq.\ (\ref{Gamma}); 
the curve labeled ``momentum conservation solution" shows the analytic
form for a fully radiative blast wave \citep{bm76,cd99}. The inset
gives  $P$ as a function of observer time using the numerical solution to eqs.\ (\ref{dPdx}), 
(\ref{U(x)1}), and (\ref{dUadidx1}),
for different values of $\zeta$.

Fig.\ 4b shows the terms $P\Gamma(dm/dx)$ and $-(P+P^{-1})(d\up/dx)_{adi}$
 in the numerator of eq.\ (\ref{dPdx}) 
for Parameter Set 1 with $\xi = 0, 0.5, 0.9$, and 0.99, and 
Fig.\ 4c shows the terms $M_0$, $m(x)$ and $\up(x)$ in the denominator
of eq.\ (\ref{dPdx}). A reasonable approximation to 
 blast-wave dynamics can be obtained, when the blast wave is relativistic, by neglecting 
the $(d\up/dx)_{adi}$ term because  this term is a small, constant fraction of 
the sweep-up term $P\Gamma(dm/dx)$.
When the blast wave becomes nonrelativistic, at $x\cong 4\times 10^{17}$ cm 
for these parameters, the $(d\up/dx)_{adi}$ term
must be retained. 
In Fig.\ 4c, the range of $x$ where the $M_0$, $\up(x)$ amd 
$m(x)$ terms dominate the value of the denominator of eq.\ (\ref{dPdx}) 
define the coasting, relativistic
self-similar, and non-relativistic self-similar regimes, respectively. 
In the Case 1 scenario, after the 
instantaneous discharge at $x_0$, the blast wave rapidly decelerates, 
but then recovers its adiabatic behavior as it 
sweeps up additional material from
the circumburst medium.

An instantaneous discharge may be oversimplified, however, because a discharge of any 
sort must take place over  a finite time exceeding at least $\Delta r^\prime/c$, the
light travel timescale across the blastwave width. For this reason we now consider the
Case 2 Scenario, again with Parameter Set 1.  Here the blastwave is assumed to recover its adiabatic 
behavior at $x_0 = 5\times 10^{16}$ cm after suffering an exponential discharge with
comoving (exponential-)loss timescales $t^\prime_{esc} = 10$, 20, 40, and 80 ks.
Although this is sufficient to solve the blastwave dynamics, at least in the limit $\e_{e-1}\lesssim 1$,
it  still is not clear what is the shape of the light curve.

We can derive some {\it idealized light curves} by weighting internal energies and rates
by the beaming factor $\Gamma^2$ that boosts energy and rate in a spherical blastwave. The following
weightings are employed, keeping in mind that this sort of approach neglects particle cooling 
and spectral effects. The weightings considered are 
\begin{enumerate}
\item $\Gamma^2 \up$, so that the comoving power is assumed to be proportional to the total 
internal energy, as might correspond to a single species emitter that radiates in the adiabatic limit;
\item $\Gamma^2 L^\prime$, where the comoving power $L^\prime  $ 
is set equal to the swept-up power $ 4\pi x^2 n_0 m_pc^3 \beta(\Gamma^2 -\Gamma )$, 
so this case  would represent the bolometric leptonic luminosity of a GRB radiating in the
strong cooling regime;
\item $\Gamma^2 U^{\prime 2}$, so that the comoving power is proportional to the square of the 
internal energy, as might hold in a scenario involving wave-particle coupling; and 
\item $\Gamma^2 U^{\prime 2}/(x/x_d)^3 $, where the comoving power is assumed proportional to the product
of the total internal energy and energy density, for example, of the magnetic field.
\end{enumerate}

Fig.\ 5 shows some numerical solutions for the Case 2, Parameter Set 1 Scenario
with the different weightings just described. The blast wave is assumed to 
recover its adiabatic character at $x_0/x_d \cong 4.2$. The notation here
is that $\up = 10^\#\up_\# $ ergs and $L^\prime = 10^\# L^\prime_\#$ ergs 
s$^{-1}$. Overlooking the shape of the light curves at early times---a point to which we 
return soon---these  
synthetic light curves resemble the Swift BAT/XRT X-ray light curves of GRBs. On top 
of these generic light curve shapes are the X-ray flares made by some as yet unspecified
mechanism.

%\clearpage
\begin{figure}[t]
%\vskip-0.5in
\epsscale{0.6}
\plotone{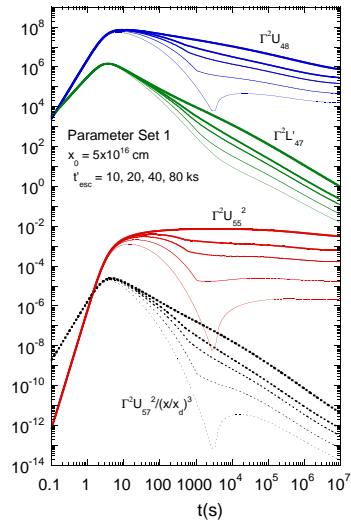}
%\vskip-0.8in
\caption{Idealized light curves for the Case 2, Parameter Set 1 Scenario with 
$x_0 = 5\times 10^{16}$ cm and $t^\prime_{esc} = 10$, 20, 40, and 80 ks.
 }
%\vskip0.1in
\label{f5}
\end{figure}
%\clearpage

The various weightings reflect different underlying assumptions of the 
physical model and yield a variety of temporal behaviors that could explain the
range of X-ray light curves observed with Swift.  The sensitivity of the 
model light curves to the location $x_0$ where the blast wave begins
to evolve adiabatically and the comoving discharge timescales is illustrated
in Fig.\ 6, showing model light curves with a range of X-ray declines and plateau-like
features.

%\clearpage
\begin{figure}[t]
%\vskip-0.5in
\epsscale{1.0}
\plotone{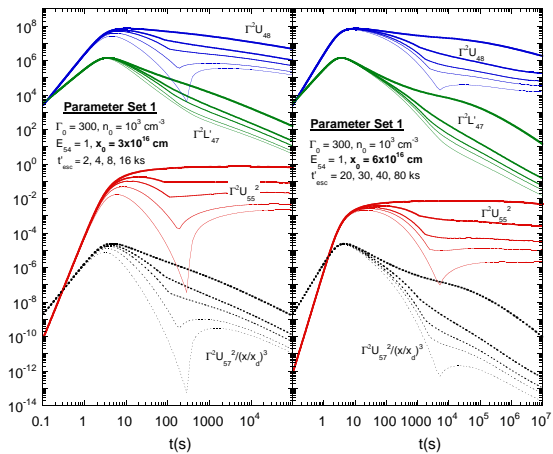}
%\vskip-0.8in
\caption{Idealized light curves for the Case 2, Parameter Set 1 Scenario with 
(a) $x_0 = 3\times 10^{16}$ cm and $t^\prime_{esc} = 2$, 4, 8, and 16 ks, and
(b) $x_0 = 6\times 10^{16}$ cm and $t^\prime_{esc} = 20$, 30, 40, and 80 ks.
 }
%\vskip0.1in
\label{f6}
\end{figure}
%\clearpage

\section{Discussion}

The central idea proposed here is that 
the rapid X-ray declines and plateaus discovered  in $\sim$ keV light curves of 
GRBs by the Swift team \citep{bar05,tag05,obr06} are signatures of UHECR acceleration in GRBs. The solution to 
the problem of UHECR origin may involve a variety of source classes, but 
must involve at least one source class, and
GRBs offer a very attractive solution:
They are found outside our Galaxy (and possibly in our Galaxy; see
\citet{wda04,mel04,dh05}), which must be the case to make UHECRs with Larmor radii larger than can be 
contained in the Milky Way; they are energetic, with each explosion releasing  as much as 
$\sim 10^{52}$
ergs kinetic energy \citep{fb05,ld07}, much of it in very clean, highly relativistic bulk particle or field-dominated 
outflows; and
they are impulsive, at least on recurrence times $\gtrsim $ day. 
This last is important by leading to a prediction of 
no clustering or angular correlations on the sky of UHECRs originating from GRBs
\citep{wm96}.

Simple (pre-BeppoSAX) BATSE estimates for the required local emissivity in 
super-GZK $(\gtrsim 10^{20}$ eV) UHECRs compared with the emissivity in X-rays and $\gamma$ rays 
from GRBs, assumed to be at a mean redshift $\langle z\rangle \approx 1$, agree to within $\sim 1$ order
of magnitude, suggesting a connection between the two \citep{wax95,vie95,der02}. 
If GRBs are the progenitors of UHECRs, then
the superposed cosmic-ray intensity spectrum from all the UHECR GRB
sources over cosmic time display features of propagation. Most
pronounced are the GZK feature and the ankle feature, the latter of which is likely due to
pair production \citep{bg88,wda04,bgg05}.  If Auger measures a spectrum similar
to the HighRes spectrum at energies $\gtrsim 10^{19.5}$ eV, then 
these spectral features of cosmic rays, accumulated during transport from the 
source to the detector, are reasonably well-explained within a scenario where UHECRs originate
from cosmic rays, with clearcut predictions for GZK neutrino fluxes.

Here we extend the scenario that UHECRs originate from GRBs by entertaining the hypothesis
that the X-ray light curves of GRBs show a spectral feature associated with UHECR
acceleration. The synthetic bolometric
light curves in Figs.\ 5 and 6 are idealized, but contain the basic features of more realistic calculations
that depend on a wide range of underlying assumptions about particle acceleration and the fraction of energy 
going into magnetic fields and waves. Depending on the radiation model, the internal and 
external shock contribution, and the setting of the zero of time, a wide range of X-ray 
decays can be understood within this picture. 

An important feature of this scenario is that a large fraction of the 
swept-up internal energy resides in protons accelerated to ultra-high energies. 
Photopion
processes effectively discharge the particle energy in the form of a neutral beam \citep{ad03,da04}. 
The decay of the internal particle
energy density weakens the magnetic field through feedback of the particle energy 
into field energy, halting further acceleration while hastening escape
of the remaining UHECRs found in the blast wave, until a dominant fraction, $\gtrsim 90$\% 
(depending on the initial amount of energy contained in the nonthermal protons), is lost
from the blast wave as a neutral beam.

Developing detailed physical models to simulate leptonic and hadronic acceleration and loss
is beyond the scope of a single paper, but the various physics
issues that enter into a combined leptonic/hadronic GRB model can be described in somewhat 
more detail.
We break the discussion into a consideration of 
\begin{enumerate}
\item the acceleration mechanism; 
\item light curves in an internal/external shock
scenario; 
\item light curves in an external shock scenario; 
\item explanation for the rapid X-ray declines;
\item explanation for the plateau phase; and
\item predictions of this scenario.
\end{enumerate}

\subsection{Acceleration}

Acceleration to super-GZK energies is possible via 
stochastic processes in the blast-wave shell 
\citep{dh01} and through acceleration by the internal shocks in a colliding shell model \citep[e.g.,][]{wax95}. 
Acceleration to such energies through shock Fermi processes is not possible 
through relativistic external shocks formed in a surrounding medium with magnetic field $\sim \mu$G \citep{gal99},
but might be possible if the GRB takes place
 in a highly magnetized environment, as might be expected if
the surroundings are formed by the 
stellar winds of high-mass stars \citep[e.g.,][noting limitations imposed by total energy 
contained in the magnetized wind]{vb88}, or if the upstream field is amplified
by streaming cosmic rays, for example, by the \citet{bl01} mechanism \citep{dru03}.

Here, 
acceleration of swept-up protons to ultra-high energies is assumed to
take place by second-order gyroresonant processes in the blast wave shell. 
This is reflected in the derivation of the acceleration rate, 
eq.\ (\ref{tp_1acc1}),
which contains only shocked fluid quantities. 
A turbulent, stochastic Type 2 Fermi mechanism in the blast-wave shell
formed by a relativistic (internal or external) shock can make 
a highly efficient accelerator \citep{dh01}, which is a sort of turbulent boiler
discussed decades ago \citep[][and references therein]{op75}, though here found in the shocked fluid
shell of a relativistic blast wave.
The basic coupling involves  gyroresonant acceleration of ions and electrons 
interacting with MHD wave turbulence \citep[in the context of Solar flares, see, e.g.,][]{sm92,mr89,mr95}, 
the form of which is model-dependent and 
 may furthermore involve anisotropic coupling depending on wave type \citep{gol01}.  
But for isotropic power-law wave turbulence,
acceleration through the stochastic Fermi mechanism makes
hard number spectra ($n(\gamma)\propto \gamma^{-s}$), with $s \approx 1$ and most of the
energy content consequently in the highest particle energies \citep{sch84,sch89,dml96,dh01}. 
Combined first- and second-order processes \citep[e.g.,][]{sch84,kru92,os93} suggest that steeper
spectra, with $1.5\lesssim s \lesssim 2$, would be formed in this sort of scenario. 
But if $s \lesssim 2$, most of the energy resides in the highest energy particles, which 
can  be accelerated to energies $\gtrsim 10^{20}$ eV \citep{der06}. 

The gyroresonant wave-particle interactions accelerate particles dynamically,
so no steady state is reached \citep[analytic solutions to time-dependent particle evolution
through second-order processes are given by, e.g.,][]{pp95,bld06}, 
and energy flows between waves, particles, and fields
until it is discharged from the system. The bulk kinetic energy of the blast wave,
initially dissipated in the shell as  highly nonthermal internal particle
kinetic energy as well as field and wave energy, is transformed into a component of 
ultra-relativistic protons carrying a large fraction of the total energy.
This internal energy, including the wave energy that continues 
to be fed into the particles, is discharged when photopion 
losses become sufficiently great. Growth of the target
photon field, for example, from proton synchrotron and photohadronic
secondaries followed by electromagnetic cascading, can lead to an 
photohadronic or proton-synchrotron loop instability of the type considered by 
\citet{km92} to extract increasingly more energy of the UHECRs until the internal
energy content is effective discharged. Obviously, more model studies are required to 
quantify the total fraction of internal energy that can be discharged in this way.

Under the given circumstances, $\sim 10^{18}$ eV photopion neutrinos are created from the 
decay of photomeson secondaries, 
ultra-high energy neutrons escape if they avoid further photopion 
interactions, and an electromagnetic channel consisting of 
$\gamma$ rays and e$^+$ and e$^-$ initiates a cascade,
ultimately leading to a flux of escaping $\gamma$ rays \citep{ad03}. 
The cosmic-ray neutron discharge from GRB blast
waves, as considered here, produces a component of the UHECRs, causing the 
ionic composition of UHECRs to be 
proton-dominated.  The decaying UHE neutrons and, indeed, photoprocesses by the highest 
energy protons and ions make an UHE synchrotron decay halo \citep{der02,ikm04} around host galaxies. 
The mean injection spectrum per GRB by this process has not
been calculated from first principles, but could range from a flat $\gamma^2 \dot n(\gamma)$ over
a narrow energy range $\sim 10^{19}$ -- $10^{21}$ eV, as in the model of \citet{wb99}, 
to a power-law injection $\gamma^2 \dot n(\gamma)\propto\gamma^{-0.2}$, as in
the model by \citet{wda04}. In principle, this process could also operate in 
radio/$\gamma$-loud AGNs, and produce spectra as steep as $\gamma^2 \dot n(\gamma)\propto\gamma^{-0.7}$ 
found in the models of \citet{bgg06}.

The acceleration mechanism in a colliding shell scenario
is much different than stochastic acceleration in the fluid shell,
 but the photohadronic discharge mechanisom
could certainly operate in such a system, though the details
still need to be worked out. Incidentally, the potential importance of 
second-order processes in GRB blast waves may 
negate the concerns
of \citet{gcl00} that the GRB must display a strong-cooling spectrum. Second-order
acceleration competing with leptonic radiative and adiabatic energy losses can form a low-energy pileup
in the electron distribution, so that no cooling tail appears.  Indeed, 
the shape of broadband time-averaged spectra of GRBs \citep{sch98}, interpreted as 
nonthermal leptonic synchrotron radiation, seems to require 
an abrupt low-energy cutoff and may be consistent with an electron pileup at some lower energy due
to combined first- and second-order processes with cooling.
This question will take greater urgency when GLAST provides combined GLAST GBM, GLAST LAT, and
Swift BAT and XRT GRB light curves, permitting detailed spectral analysis of bright GRBs 
from the prompt to afterglow phases. This will reveal the transition from the 
internal shocks (in the internal/external shock scenario) to external shocks, as now considered.

\subsection{Internal/External Model}

In the internal/external scenario \citep{pir99,pir05,mes06}, the transition from the prompt phase 
formed by colliding shells of jet plasma to the afterglow phase made by 
external shocks takes place no later than 
the time of BeppoSAX reorientation, $\sim 8$ hrs, and likely much sooner.
In this picture, the internal shocks make a distinct radiation signature from
the external shock emission received later.
The zero of time for the external shock emissions from the  GRB can, in the internal/external
scenario,
be set to various times, but is determined observationally  when the GRB detector is triggered, e.g.,
by precursor emission.  In the case of a GRB exploding in a uniform circumburst 
medium, the latest the zero of time can be set would be about when the external 
shock signature reaches its maximum brightness and triggers the GRB detector, 
which is generally near the deceleration
time $t_d$. The freedom to set the zero of time \citep{lb06,kz07} when modeling data with 
 the external shock component
considered here, as shown in Fig.\ 7,
adds an additional distortion and increased variety
to the model light curves, similar to various behaviors of 
the X-ray light curves found with Swift.

%\clearpage
\begin{figure}[t]
\epsscale{0.6}
\plotone{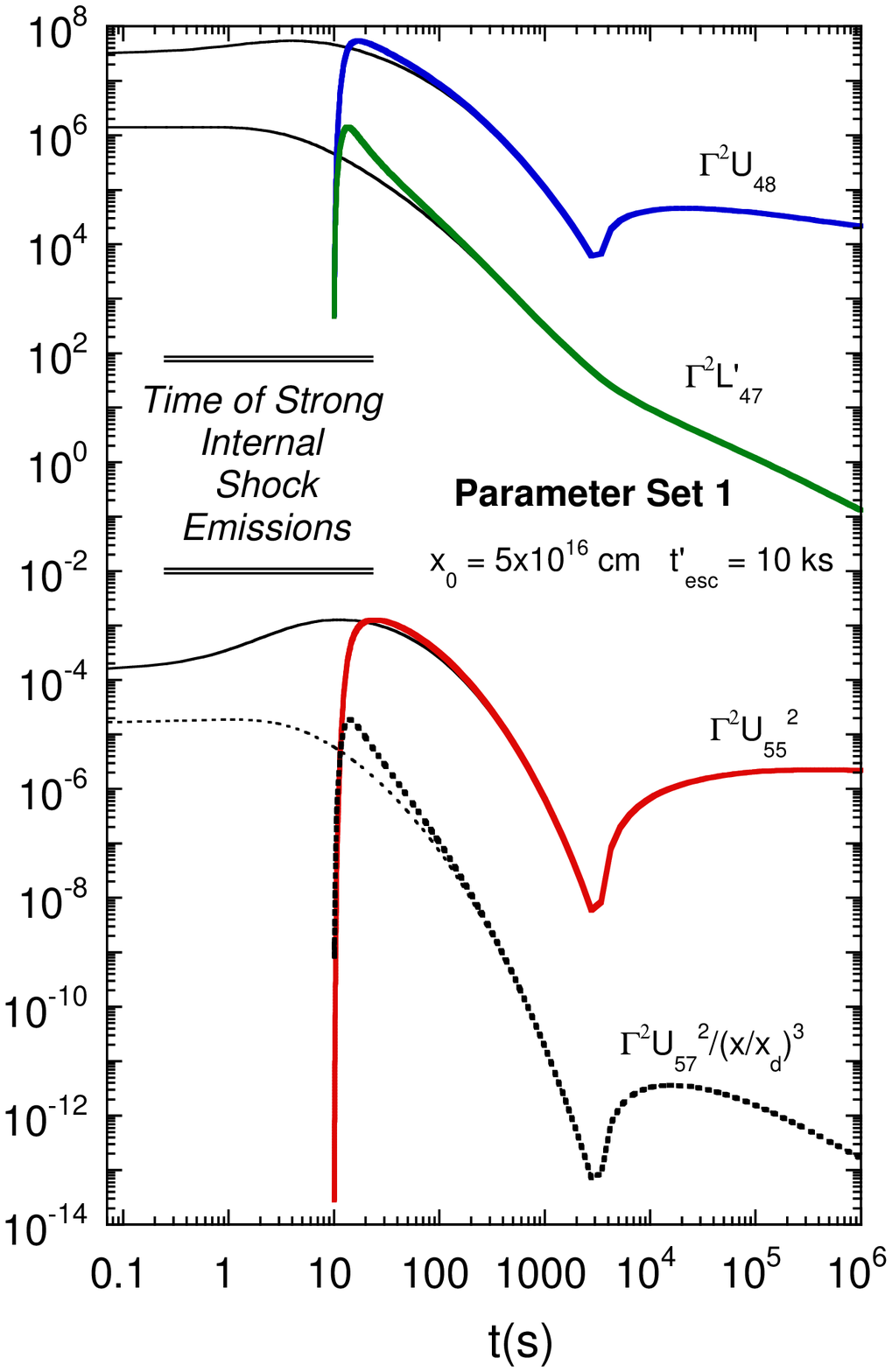}
\caption{Idealized light curves for the Case 2, Parameter Set 1 Scenario with 
 $x_0 = 5\times 10^{16}$ cm and $t^\prime_{esc} = 10$ ks, with the time
reset to -3 s and + 10 s for the thin and thick curves, respectively. }
\label{f7}
\end{figure}
%\clearpage

An alternate view of the internal/external model is to have all the prompt emissions
up to the time of the rapid X-ray declines made by internal shock processes (see \S 5.5). 
Here it is surprising that the Swift BAT light curves are rather smooth and 
apparently not a superposition of generic kinematic curvature light pulses \citep{yam06,der04}, 
although the last pulse in some GRBs, for example, Swift XRT data of GRB 061202 $\sim 120$ s
after burst trigger \citep{gcn19}, do appear possibly to represent the last major accretion
events from a hypothetical collapsar torus.
In either variant of the internal/external scenario,
 a complete model GRB light curve would combine the 
model light curves calculated in an external shock treatment with a 
second component consisting of $\gamma$-ray pulses
and X-ray flares from colliding shells made by central engine activity. 

Placing these calculations in the context of the internal/external approach, 
the transition from internal shock to 
external shock emissions takes place, typically, between 50 and 100 seconds, 
precisely in the gap between BAT and XRT, at which time
the flux levels of the internal and external shock components are roughly equal.
It may seem coincidental that  the times for the 
 transitions from internal to external shocks are 
$\approx t_d$ \citep[cf.][]{zha06}. Swift observations of late-time X-ray flares 
interpreted as central engine activity mean, however, that the internal shocks  
operate over a wide range of times, nullifying to some extent this coincidence.
It may also appear coincidental that the flux level of the 
rising external-shock component is at about the 
same level as the earlier prompt internal shock component. 
But the large fraction ($\gtrsim 90$\%) of swept-up energy in the form of 
nonthermal hadrons, the factor $10$ -- 100 more nonthermal energy in the shells before the 
radiative discharge compared to after, and the low efficiency $\sim 1$ -- 10\% for 
transforming internal shock energy to 
hard X-rays and $\gamma$ rays \citep{bel00,kum99,gpw05}, naturally leads to the expectation that 
 energy radiated
in afterglow phase is comparable to that of the prompt phase, so that the similar flux levels of the 
internal and external shock components may  not be so surprising.

We conclude that the proposed interpretation for the fast decays is
in accord with the internal/external shock scenario for GRBs, provided that the external 
shock emissions make a significant contribution already in the early afterglow. 

\subsection{External Shock Model}

A conceptually simpler though not generally accepted approach is to suppose that 
all GRB emissions are formed by a single impulsive event. In such a 
picture, the $\gamma$-ray pulses and X-ray flares
seen in GRB light curves are a consequence of interactions of the 
blast wave with a clumpy circumburst
density \citep{dm99,dm04}, and there is a single blast
wave that is energized only by the external forward/reverse shock pair.

The central requirement for short timescale variability in this model is 
that the size scale of the inhomogeneities (clouds) is $\ll x/\Gamma_0$ \citep{dm99},
and that the blast wave remains thin in order to form a strong forward shock, 
which translates into the condition that the blast wave spreading is $\lesssim x/\Gamma_0^3$
\citep{der06,der07}. Even late-time X-ray flares could be the result of 
a GRB blast wave intercepting clumps of material if the blast-wave shell remains thin \citep{der07}. 
A prediction of this model is that ranges of baryon loading, in particular, mass-loaded
dirty fireballs \citep{dcb99}, can account for the differences between classical 
long duration GRBs and X-ray flashes.

This model has been questioned on ground of efficiency \citep{sp97,pir05}, but the argument
is incorrect \citep{dm04}. 
The additional energy content in the prompt phase makes high efficiency 
for the external shock model, as for the internal shell model,
less critical, improving the viability of both models. Another concern is that
late sharp flaring events cannot be made by external shocks. The simulation 
of  \citet{zha06} treats however a large cloud, and the treatment
of \citet{ng06} operates in the adiabatic self-similar phase, after which much of the 
blast-wave energy has already been converted to radiation \citep[this mistake, and the 
failure to take into account beaming factors for the interaction,
is also made by ][]{iok05}. The brightest flares 
are made by portions of the blast wave that have not yet entered this phase, 
and which intercept a dense clump of material with a thick column \citep{dm99}.  See \citet{der07}
for a fuller discussion. 

Whether or not a ``pure" external shock model for GRBs can be defended,
the proposed model is already in accord with the internal/external
scenario, as already described. It is also interesting to note that
\citet{mkp04} find that the afterglow $\e_B$ parameter must be much weaker
than in the prompt phase to explain light curves of certain smoothly varying GRB.

\subsection{Explanation for the Rapid Decay}

We have considered a physical mechanism to explain the rapid decays
observed with Swift from GRBs in the late prompt/early afterglow phase that 
could operate in both internal/external and external shock scenarios. The number of 
physical models for this effect is not yet large;  there is a preference recently 
\citep[e.g.,][]{nav07,pan06,obr06a,zlz06,lia06}
to quantify the effect and perform empirical phenomenological studies 
rather than to construct physical models. 
An exception is \citet{pmr06}, who consider scattering of GRB emission 
in a cocoon formed as the blast wave jet emerges from the stellar 
photosphere \citep[see also][]{pmr06a}. The emergent photons can exhibit a steep temporal decay, 
similar to the  GRB X-ray light curves observed with Swift XRT. This model applies strictly to a collapsar
scenario, where the shocks emerge from the stellar photosphere by forming a cocoon \citep[e.g.,][]{rcr02},
yet rapid decays also occur in short, hard GRBs \citep[e.g., GRB 050724,][]{bar05a}, which are thought
to originate from coalescence events that lack stellar photospheres and cocoons 
\citep[for recent reviews of short hard GRBs, see][]{nak07,lrr07}.\footnote{This invites the 
speculation that the short hard class of GRBs are also 
sources of UHECRs, given appropriate parameters, in particular, small variability times.}

The curvature effect \citep{kp00,der04,lia06}, which limits the rate at which an impulsively
illuminated shell can temporally decay, obviously plays an important role. The curvature limits
apply to the case of an active central engine that suddenly shuts itself off, or dramatically
reduces its activity, as in the model of \citet{pz06}. For a spherical blast wave that is roughly 
uniform within the Doppler cone, the curvature limits would also apply to internal or external shocks.
The model presented here
is in accord with the curvature limit, keeping in mind the uncertainty in 
setting the zero of time \citep[see also][]{lb06}.

From a comparison of the idealized bolometric light curves in Figs.\ 5 --8 
with the Swift BAT and XRT
GRB light curves shown in, e.g.,  the \citet{obr06} paper, it seems that
the model offers adequate variety to explain the rapid X-ray decays, and not 
 simply an average light curve \citep{obr06a}.
% that camouflages the wide range
%of light curve shapes and characterizes almost no actual GRB.

\subsection{Explanation for the Plateau Phase}

Focusing on the external shock component in the proposed scenario,
about the time that Fermi processes
can accelerate a large fraction of the particle energy to ultra-high energies,
photohadronic losses and escape become important and 
cause a sudden loss of internal energy. The magnetic field energy, which is supported 
by the internal energy, itself becomes smaller,  thereby 
enhancing nonthermal particle escape and causing the internal energy to collapse. 
At some point, the energy content in 
the blast wave reaches its minimum when only the original baryon load and a remnant
magnetic field remains. Magnetic field generation processes like the Weibel instability \citep[e.g.,][]{nis06}
could then amplify the blast wave magnetic field until  plasma is swept-up and
entrained, and the blast wave again commences to evolve adiabatically.

The viability of this explanation for the plateau phase and the diversity of X-ray light curves
discovered by Swift \citep{bar05} is suggested by the wide range of idealized light 
curve profiles calculated and plotted in Figs.\ 5 -- 7, though more research will be required
to quantify the underlying assumptions
about the acceleration and particle radiation mechanisms that will allow fits to data. 
This explanation is quite different from 
the conventional explanation 
in the internal/external scenario. There, delayed energy injection  \citep[e.g.,][]{zm01,zha06}, possibly
signaled by the late-time X-ray flares,  could harden the temporal decay. 
Yet plateau phases are often seen soon after a rapid decay, as in the 
cases cited in the Introduction, so 
the central engine activity has to be concealed while powering the light 
curve. 
An analysis of GRB 050713A by \citet{gue07} starts with the conclusion that 
the behavior of GRB 050713A requires refreshed shocks.
\citet{tom06} consider a patchy jet scenario where the delayed off-axis emissions
become brighter than the on-axis emissions, but their model lacks a self-Compton 
component, and the afterglow X-ray component lasts much longer than found in detailed
numerical calculations with similar parameters \citep{dcm00}.
\citet{fp06} try to model this behavior and assess
the energy requirements.  They claim that powering
the hard afterglow phase does not make excessive demands on energetics, but 
do not display blast-wave solutions characteristic of the rapid decay and plateau 
phases discovered \citep{nou06,zha06} with Swift.

\subsection{Predictions}

Besides the steep decays and plateau phases in the X-ray light
curves from GRBs, there are other direct radiative signatures for the hypothesis that
UHECRs originate from GRBs.  In all cases where photopion losses are important, 
the blast wave is assumed to be radiating in
the fast-cooling limit, so that the $\nu F_\nu$ indices $a \cong 1/2$
and $b = 1 - (s/2)$, where $s$ is the electron injection index (with
$s\approx 2.2$ for shock Fermi models). If we accept that accelerated 
electrons radiatively cool without being subsequently accelerated (contrary,
incidentally, to the point made earlier with respect to the concerns 
of \citet{gcl00}),
then already a careful spectral
analysis for the variation of $\e_{pk}$ predicted by external shock 
emissions as it sweeps through the Swift XRT detector could
see if GRB blast waves evolve within a few hundred seconds to a
condition where the $\e_e$ and $\e_B$ parameters are each $\gtrsim
10$\%.

A prediction of this model is hadronic $\gamma$-ray light consisting
of proton synchrotron, photopion and secondary photohadronic decay
radiations that cascade to low enough energies that internal and diffuse cosmic $\gamma\gamma$
 attenuation processes become negligible.  This component varies
differently from the X-ray lepton synchrotron component and can appear
much after the prompt emissions. The hadronic emission component could then 
possibly account for the delayed $\gamma$-ray emission in
GRB 940217 \citep{hur94} and GRB 941017 \citep{gon03}, likely in addition
to SSC emission. The independently varying hadronic emission component is a
prediction for GLAST, though a leptonic model 
\citep[e.g.,][]{pw04}
for delayed anomalous components that includes Compton-scattered forward- and
reverse-shock emissions, could also make a delayed 
component as observed from GRB 941017.

Ultra-high energy neutrinos are made by the release of blast wave
energy though this mechanism, and we predict the detection of a
delayed neutrino flux at high energies just about the time of the 
rapid decays. Unfortunately, the $\gtrsim
10^{17}$ eV neutrinos from the photohadronic processes in GRB blast
waves are at a difficult energy for IceCube to detect, but the
$\gtrsim 10^{18}$ eV neutrinos could be detected in sideways showers
with Auger, or in balloon-borne detectors such as ANITA.  
%The prediction of the PeV neutrino flux for external shock processes in
%the prompt phase have only been made for a uniform surroundings, where
%the rate is too low to be detectable with IceCube \citep{der02}, though a
%clumpy medium may improve detectability.  
Besides the UHE neutrino component, PeV neutrinos detectable with IceCube will be made during the prompt phase
 in the internal shell/collapsar model in large fluence GRBs when the shocked shell Lorentz
factors $\sim 100$ -- 200, while being optically thick to $\gamma\gamma$
attenuation at GeV energies \citep{da03,rmz04,gue04}.

For parameters that make a highly
radiative blast wave phase, $\e_{p,syn}$ for proton synchrotron
emissions from UHECRs in GRBs will be found in the TeV range. To avoid
attenuation by the diffuse intergalactic infrared radiation fields,
this emission could only be detected from
low redshift, $z \lesssim 0.4$, GRBs, which 
can be measured with low-threshold air Cherenkov telescopes such as
MAGIC. The photon flux at these energies is very low, so only a few
$\gtrsim 100$ GeV photons could be detected with a direct photon detection
telescope such as GLAST.

\section{Summary and Conclusions}

By carrying through an analysis of a complete leptonic/hadronic GRB
blast-wave model, a region of parameter space was found for GRB 
external shocks where, even in
the simplest constant density blastwave model, photohadronic
processes have important effects on blastwave evolution.  
This occurs for the observer a few hundred seconds after the start of the
GRB for circumburst medium densities $n_0\gtrsim 10^2$ cm$^{-3}$, $100
\lesssim \Gamma_0 \lesssim 300$, $E_{54} \sim 1$ -- 10, and
$\e_e,\e_B\gtrsim 10$\%.  For these parameters, 
the time scales for photopion losses and
direct escape become comparable to the available time in the early 
afterglow, with observable
consequences.

The GRB blast wave is predicted to evolve toward the strong cooling
regime in the early afterglow, leading to a strong discharge
of energy through photohadronic processes 
and direct escape of UHECR ions, causing the declines in the X-ray 
light curves observed by Swift.  The change of $\e_{pk}$
with time may indicate the evolution of the radiative regime during
the prompt and early afterglow phases, though this component can 
be concealed in an internal/external shock model by internal 
shock emissions.  Hadronic
radiations consisting of high-energy neutrinos, cascade $\gamma$ rays,
and escaping cosmic-ray neutrons could be important to
$\gtrsim 10^{4}$ s after the GRB, producing a component that varies, 
in general, differently from the X-ray lepton sychrotron emission.
As a prediction for GLAST, this component should be
bright at GeV energies just before the times that the rapid X-ray declines are
observed. A related flux of $\gtrsim 10^{17}$ eV neutrinos is made at
this time.

The release of energy in the form of neutrons or escaping ions in the 
early afterglow phase means
that there was much more energy in the GRB blast wave during the 
prompt phase. Thus the efficiency of $\gamma$-ray production during 
the prompt phase is low, which ameliorates efficiency concerns
in both external or internal shock models.
Hadronic processes imprint GRB light curves and the UHECR spectrum
with distinctive signatures. The rapid X-ray declines and plateau
phases discovered with Swift are argued here to be just such signatures.

\acknowledgments
I thank A.\ Atoyan, M.\ B\"ottcher, and J. Chiang for
collaboration and comments, and Dr.\ E.\ Nakar for 
pointing out how this solution could resolve efficiency concerns
in GRB models. I also sincerely thank the referee for a thorough
and constructive report. This work is supported by the Office of Naval Research,
by NASA {\it GLAST} Science Investigation No.\ DPR-S-1563-Y, and 
NASA Swift Guest Investigator Grant No.\ DPR-NNG05ED41I.

\end{document}